\documentstyle[prl,twocolumn,aps]{revtex}
\begin{document}

\twocolumn[\hsize\textwidth\columnwidth\hsize\csname
@twocolumnfalse\endcsname
\title{Persistent Currents in Charge-density Wave Systems}
\author{N.~Harrison
}
\address{
National High Magnetic Field Laboratory, LANL, MS-E536, Los
Alamos, New Mexico 87545
}
\date{\today}
\maketitle

\begin{abstract}
The inductive exchange of carriers between closed Fermi surface
sections subject to Landau quantization and open Fermi surface
sections subject to charge-density wave (or spin-density wave) formation
is shown to give rise to persistent currents. This
mechanism may explain recent experimental data in certain organic conductors
in high magnetic fields.
\end{abstract}

]\narrowtext

It is well known that a gradient in the orbital magnetization within a
bulk metal corresponds to a net effective electrical current density
${\bf j}=\nabla\times{\bf M}$~\cite{bleaney1}.  Such currents can be
true macroscopic flows of charged quasiparticles, created in response
to changes of magnetic field ({\it i.e.} $\partial {\bf H}/\partial
t$) and, in the case of superconductors, applied magnetic field ${\bf
H}$.  They are considered to be ``persistent'' if they exist
in equilibrium, and truly persistent currents are understood to exist
only in superconductors~\cite{tinkham1}.  In the type I
superconducting phase, the Meissner effect screens all electromagnetic
fields from within the bulk~\cite{tinkham1}, while in the type II
superconducting phase, net currents that permeate the bulk occur
transverse to a vortex density gradient held in place by vortex
pinning~\cite{tinkham1}.  Long-lived currents~\cite{jones1} also occur
in quantum Hall systems orthogonal to a chemical potential gradient
$\nabla\mu$~\cite{chakraborty1}.  The equilibrium Hall electric field
is sustained owing only to the absence of quasiparticle scattering
processes orthogonal to the current~\cite{chakraborty1}.  

In this
paper, I identify a further mechanism for persistent currents
involving the coexistence of orbital magnetism with a charge-density
wave (CDW) or spin-density wave (SDW) modulation~\cite{gruner1}.
With the pinning of a CDW (or SDW) modulation~\cite{gruner1}, a
situation can be realised whereby a differential chemical potential
gradient $2\nabla(\Delta\mu)$ exists between the CDW (or SDW) and
additional {\it ungapped} sections of Fermi surface present.  Currents
result within the bulk if, and only if, the derivative $\partial{\bf
M}/\partial(\Delta\mu)$ is finite within the bulk.  This quantity is
vanishingly small in most metals at low magnetic
fields~\cite{reason1}, but can become quite significant when Landau
levels are formed at high magnetic fields~\cite{maniv1}.  The
development of the gradient $2\nabla(\Delta\mu)$ relies on there being
a continual relative exchange of carriers $2\partial(\Delta
N)/\partial|{\bf H}|$ between Fermi-surface sheets as the magnetic
field is swept~\cite{maniv1}.  Again, this is a negligible effect at
low magnetic fields, but becomes very significant when Landau levels
are formed at high magnetic fields~\cite{maniv1}.  The mechanism for
persistent currents in CDW (or SDW) systems, described here, will
therefore be seen to be quite different than that predicted by
Fr\"{o}hlich~\cite{frohlich1}.

A multiband metal consisting of separate one-dimensional (1D) and
two-dimensional (2D) Fermi surface sections is an ideal system in
which to model these effects.  Fermi surfaces of this topology are
found to occur in several organic metals~\cite{mori1,neilbs}, and
effects consistent with the existence of persistent currents have been
observed experimentally~\cite{harrison1,harrison2,harrison3,ardavan1}.

I begin by considering a metal that consists of only a single 2D
pocket, characterized by a de Haas-van Alphen (dHvA) frequency $F$ and
cyclotron frequency $\omega_{\rm c}=eB/m^\ast$, where $B={\bf
B}\cdot\hat{\bf z}$.  When the quasiparticles move in planes
perpendicular to the unit vector $\hat{\bf z}$, Landau levels with
eigenvalues $\varepsilon=\hbar\omega_{\rm c}\big(n+\frac{1}{2}\big)$
are formed, where $n =0,~1,~2,\dots$.  The chemical potential $\mu$ is
pinned to the highest occupied of these in order to conserve the total
number of 2D carriers, $N_{\rm 2D}$~\cite{shoenberg1}.  It is
instructive to consider contrasting situations at half-integral and
integral Landau level fillng factors $\nu=F/B$; at half-integral
filling factors (when $\nu+\frac{1}{2}$ is an integer) the highest
occupied Landau level is exactly half filled, while at integral
filling factors (when $\nu$ is an integer) $\mu$ is in the
transitional region between filled and empty levels.  In the low
temperature limit ($T\rightarrow$~0), the corresponding
susceptibility components $\chi_{zz}=\partial M/\partial H$
are~\cite{shoenberg1}
\[
    \chi_{\rm half}=
    \frac{\hbar\omega_{\rm c}}{e\rho_{xy}}
    \frac{F}{\mu_0H^2}
\]
and
\begin{equation}
    \chi_{\rm int}=-\frac{\pi\omega_{\rm c}\tau}{4}
    \frac{\hbar\omega_{\rm c}}{e\rho_{xy}}
    \frac{F}{\mu_0H^2}
\label{susceptibility2D}
\end{equation}
respectively, where $\rho_{xy}$ is the off-diagonal component of the
resistivity tensor, $M={\bf M}\cdot\hat{\bf z}$ and $H={\bf
H}\cdot\hat{\bf z}$.  Note that the susceptibility at integral filling
factors is extremely sensitive to Landau level broadening that results
from a finite relaxation time $\tau$~\cite{maniv1}.  It follows that
${\bf M}$ and $\mu$ are related via the differential
\begin{equation}\label{chemp}
    \frac{\partial {\bf M}}{\partial\mu}=
    \frac{N_{\rm 2D}}{\mu_0H}~\hat{\bf z}=
    \frac{1}{e\rho_{xy}}~\hat{\bf z},
\end{equation}
where both $e$ and $\rho_{xy}$ are negative for electrons and positive
for holes~\cite{shoenberg1}.

When 1D carriers are added to the system, they primarily function as a
charge reservoir \cite{maniv1}.  At half-integral filling factors,
carriers are transferred from the 1D sheets to
the 2D pocket as $H$ is increased, causing the magnetization to
increase more quickly than in Equation (\ref{susceptibility2D}).  At
integral filling factors, the inverse situation becomes true.  The
presence of 1D states in the Landau gaps prevents ${\bf M}$ and $\mu$
from jumping abruptly between Landau levels.  The susceptibilities
therefore become
\[
    \chi_{\rm half}=
    \frac{g_{\rm 2D}}{(g_{\rm 1D}+g_{\rm 2D})}
    \frac{\hbar\omega_{\rm c}}{e\rho_{xy}}
    \frac{F}{\mu_0H^2}
\]
and
\begin{equation}
        \chi_{\rm int}=-\frac{g_{\rm 1D}}{(g_{\rm 1D}+g_{\rm 2D})}
    \frac{\hbar\omega_{\rm c}}{e\rho_{xy}}
    \frac{F}{\mu_0H^2},
    \label{susceptibility2D1D}
\end{equation}
where $g_{\rm 1D}$ and $g_{\rm 2D}$ represent the density of states of
the 1D sheets and 2D pocket respectively~\cite{maniv1}.

The process by which the 1D sheets function as a charge reservoir
becomes perturbed upon opening of a gap.  For the purpose of modeling
these perturbations, it is convenient to consider a simplified system
in which the Fermi surface nesting gaps the entire 1D density of
states and the effects of Zeeman splitting of the 1D bands are
negligible.  The latter is true in SDW systems~\cite{gruner1} and also
in CDW phases in which the spin up and spin down bands nest
independently~\cite{zanchi1,mckenzie1}.  By far the most important
consideration is that the CDW (or SDW) phase remains stable over the
field range where the dHvA oscillations occur.  Persistent currents
are anticipated only in the case of an incommensurate CDW (or SDW) in
which the nesting vector ${\bf Q}$ can adjust itself in order to
minimise the total free energy of the system.  They are not expected
in the case of a commensurate CDW phase that is pinned to the crystal
lattice~\cite{harrison4}.  Following the solution of the gap equation
obtained by Maki and Tsuneto~\cite{maki1}, adapted to include the
effects of a shift in the chemical potential (labelled here as
$\Delta\mu$) in Reference~\cite{harrison4}, the change in free energy
in forming such a CDW (or SDW) groundstate can be written as
\begin{equation}\label{freeenergy}
    F=-g_{\rm 1D}\bigg[\frac{\Delta^2_0}{2}-
    (\Delta\mu)^2\bigg]
    -F^\prime_{\rm 2D}(\Delta\mu).
\end{equation}
The first two terms of Equation (\ref{freeenergy}) are minimised when
the zero-temperature gap $2\Delta_0$ that opens on the 1D sheets is
centred about $\mu$, so that an offset $\Delta\mu$ of the gap relative
to $\mu$ costs energy.  The third term, $F^\prime_{\rm
2D}(\Delta\mu)$, accounts for an additional (either positive or
negative) change in free energy of the 2D Landau level levels incurred
by this offset.  The offset translates directly into an increase in
the number of states $\Delta N_{\rm 1D}=g_{\rm 1D}\Delta\mu$
accommodated by the 1D sheets, balanced by an equal and opposite
reduction in the number of states contained in the 2D pocket.  If $Q$
is the component of the optimum nesting vector ${\bf Q}$ parallel to
the mean Fermi velocity ${\bf v}_{\rm F}$ of the 1D sheets,
$\Delta\mu$ also results in a shift in this $Q$ given by $\Delta
Q/\Delta\mu=Qg_{\rm 1D}/N_{\rm 1D}$, where $N_{\rm 1D}$ is the
equilibrium number of 1D states.

Should $Q$ be uniform throughout the sample as a result of cooling it
in the presence of a constant magnetic field, a finite value of
$\Delta\mu$ may already be required in order to minimise Equation
(\ref{freeenergy}) owing to the $F^\prime_{\rm 2D}(\Delta\mu)$ term. 
Such an effect has already been considered to explain the suppression
of oscillations in $\mu$ observed experimentally in certain organic
metals~\cite{nam1}.  I will not consider this effect in detail for two
reasons: first, it is not expected to operate at half-integral and
integral filling factors where $\mu=F^\prime_{\rm
2D}(-\Delta\mu)=\partial F^\prime_{\rm
2D}(-\Delta\mu)/\partial\mu=$~0: second, I reveal below that it is
unlikely that a uniform $\Delta\mu$ is maintained throughout the
entire sample.

If the sample is field-cooled at $\mu=\Delta\mu=$~0, the starting
condition of the sample corresponds to the point labeled `0' in Fig. 
\ref{testloop}.  If the CDW (or SDW) is pinned, this pinning initially
prevents $Q+\Delta Q$ from changing in response to a change in the
applied magnetic field $\Delta H$.  The optimum value of $Q$ does,
however, change, leading to relative difference $\Delta Q$ between the
actual and optimal values, and finally to a shift in $\Delta\mu$. 
Since the density of states at $\mu$ in the gap on the 1D sheets is
zero, the sample begins by responding like a 2D metal in which there
are no 1D states.  From this point on, the magnetization becomes
hysteretic.  Consequently, the irreversible change in the
magnetization is characterised by an irreversible susceptibility given
by the difference between Equations (\ref{susceptibility2D}) and
(\ref{susceptibility2D1D}); hence
\begin{eqnarray}\label{chiirr}
    \chi_{{\rm irr},{\rm half}}=
    -\frac{g_{\rm 1D}}{(g_{\rm 1D}+g_{\rm 2D})}
    \frac{\hbar\omega_{\rm c}}{e\rho_{xy}}
    \frac{F}{\mu_0H^2}~,\hspace{1.5cm}\nonumber\\
    {\rm and}\hspace{7cm}\nonumber\\
    {\bf \chi}_{{\rm irr},{\rm int}}=-\bigg[\frac{\pi\omega_{\rm c}\tau}{4}
    -\frac{g_{\rm 1D}}{(g_{\rm 1D}+g_{\rm 2D})}\bigg]
    \frac{\hbar\omega_{\rm c}}{e\rho_{xy}}
    \frac{F}{\mu_0H^2}.
\end{eqnarray}
As expected for magnetic hystersis, ${\bf \chi}_{\rm irr}$ is always
negative.

For a sample without boundary conditions, the magnetization would
continue to change with the susceptibility given by Equation
(\ref{chiirr}), giving rise to the dotted line in Figure
\ref{testloop}.  This would occur until $\Delta\mu$ reaches the
thermodynamic limit, $\Delta\mu_{\rm lim}=\pm~\Delta_0/\sqrt{2}$,
whereupon Equation (\ref{freeenergy}) is no longer negative.  At this
point, the magnitude of the irreversible magnetization would saturate
at a maximum value, ${\bf M}_{\rm irr,lim}=\pm~\Delta\mu_{\rm
lim}\times\partial {\bf M}/\partial\mu=\pm~\Delta_0{\bf
z}/\sqrt{2}\rho_{xy}$, without any currents being induced within the
bulk of the sample.  Just as in superconductors~\cite{tinkham1} and
quantum Hall systems~\cite{chakraborty1}, however, a net magnetization
cannot exist within the bulk without incurring a current ${\bf j}$
immediately inside its surface.  Here, the current is carried by 2D
carriers, and wherever these exist,
\begin{equation}\label{current}
    {\bf j}=-\nabla(\Delta\mu)\times
    \frac{\partial{\bf M}}{\partial\mu},
\end{equation}
where $-\nabla(\Delta\mu)$ is the chemical potential gradient
experienced by the 2D pocket.  In order to maintain the equilibrium
condition $\nabla\mu=0$, an equal and opposite charge polarization
field $\nabla(\Delta\mu)$ must also form on the nested 1D Fermi surface
sheets.  Since CDWs (and SDWs) are electrostatically limited, rather
than increasing indefinitely, $\nabla(\Delta\mu)$ must eventually be
limited by the characteristic threshold electric field
\begin{equation}\label{threshold}
    \nabla(\Delta\mu)_{\rm lim}=e{\bf E}_{\rm t}
\end{equation}
required for its depinning.  This equates to a critical current density
\begin{equation}\label{criticalcurrent}
    {\bf j}_{\rm c}=
    \frac{{\bf E}_{\rm t}\times\hat{\bf z}}{\rho_{xy}}
\end{equation}
for the carriers in the 2D pocket within the sample.

From Equation (\ref{criticalcurrent}), one can see that the
electrodynamics has parallels with the quantum Hall
effect~\cite{chakraborty1}, with the exception that there exists zero
net electric field in equilibrium: the quantity ${\bf E}_{\rm t}$
instead parameterizes the charge polarization field required to slide
the CDW (or SDW), that ultimately leads to dissipation.  The absence
of a net electric field implies that the effect, described here, is
not limited only to integral Landau level filling factors.  The
inductive behaviour of the sample is, in contrast, much more like that
of a type II superconductor, but with an irreversible susceptibility
${\bf \chi}_{\rm irr}$ that can depart significantly from the ideal
diamagnetic value $-1$.  As $\Delta H$ increases, the
surface region in which ${\bf E}_{\rm t}$ is polarized propogates
further into the sample.  The inductive response of the sample to a
change in applied magnetic field $\Delta H$ is therefore
indistinguishable from the critical state model developed by
Bean~\cite{bean1}, provided ${\bf E}_{\rm t}$ is finite for all
directions within the planes.  This is certainly the case in CDW (or
SDW) systems for which ${\bf Q}$ occurs at an oblique angle with
respect to the lattice vectors, where Fermi surface nesting results in
charge (or spin) modulation that is two-dimensional.

If one considers a cylindrical sample of radius $r$, with $\hat{\bf z}$
aligned along its axis, following Bean~\cite{bean1}, the irreversible
magnetization becomes
\begin{equation}\label{bean}
    {\bf M}_{\rm irr}=-{\bf \chi}_{\rm irr}
    \bigg[\Delta H-\frac{(\Delta H)^2}{H^\ast}
    +\frac{(\Delta H)^3}{3H^{\ast 2}}\bigg].
\end{equation}
This region is identified by point `1' in Fig.  \ref{testloop}.  In
this model, $H^\ast$ is the characteristic coercion field while ${\bf
\chi}_{\rm irr}H^\ast$ is the maximum magnetic field that can be
screened within the sample.  When $\Delta H=H^\ast$, corresponding to
point `2' in Fig.  \ref{testloop}, the irreversible magnetization
reaches its saturation limit for a cylindrical sample,
\begin{equation}\label{saturation}
    {\bf M}_{\rm irr}=\frac{{\bf \chi}_{\rm irr}H^\ast}{3}=
    -\frac{|{\bf j}_{\rm c}r|}{3}~\hat{\bf z}.
\end{equation}
When the direction in which the field is changed is reversed at point
`3' in Fig.  \ref{testloop}, the initial susceptibility is, once
again, given by Equation (\ref{chiirr}).  A similar cubic law should
apply to that in Equation (\ref{bean}) in the reversal state, but with
double the interval in $\Delta H$ being required to reverse the
polarity of $\nabla(\Delta\mu)$.  The magnetization saturates again at
point `4,' and a full hysteresis loop results on further cycling of
$\Delta H$.

I now compare the predictions of this model with recent measurements
made on samples of the organic conductor
$\alpha$-(BEDT-TTF)$_2$KHg(SCN)$_4$ at high magnetic fields.  This
material has a Fermi surface which nearly matches that of the
model~\cite{neilbs}.  At low temperatures, $T\lesssim$~8~K, and low
magnetic fields, $\mu_0H\lesssim$~23~T,
$\alpha$-(BEDT-TTF)$_2$KHg(SCN)$_4$ is believed to possess a
commensurate CDW phase, transforming to one that is incommensurate at
high magnetic fields $\mu_0H\gtrsim$~23~T
\cite{harrison1,harrison2,harrison3,mckenzie1,harrison4,biskup1,qualls1,kartsovnik1}.  
It is within the
high magnetic field phase where effects consistent with persistent
currents have been
reported~\cite{harrison1,harrison2,harrison3,ardavan1}.

I begin my comparison by considering the irreversible
susceptibility.  On computing ${\bf \chi}_{{\rm irr}}$, using Equation
(\ref{chiirr}) and the established parameters ($F\approx$~670~T,
$m^\ast\approx$~2~$m_{\rm e}$, $N_{\rm
2D}\approx$~1.6~$\times$~10$^{26}$~m$^{-3}$, $g_{\rm 1D}\approx g_{\rm
2D}$ and $\tau\approx$~2~ps~\cite{harrison5} for
$\alpha$-(BEDT-TTF)$_2$KHg(SCN)$_4$), the model estimates ($\chi_{\rm
mod}$) listed in Table \ref{table1} are found to agree rather well
with the experimental values ($\chi_{\rm exp}$)~\cite{ardavan1} both
at half-integral and integral filling factors.

I now estimate the threshold electric field from the saturation
magnetization ${\bf M}_{\rm irr,max}$ (listed in Table \ref{table1} as
$M_{\rm max}$).  This is found to be marginally higher at
half-integral filling factors than at integral filling factors in
Reference~\cite{ardavan1}, but becomes comparable at the lowest
temperatures~\cite{harrison1}.  Making the cylindrical sample
approximation, with $r\sim$~0.5~mm, one can estimate the mean scalar
critical current $j_{\rm c}$ using Equation (\ref{saturation}) and the
corresponding threshold electric field $E_{\rm t}$ using Equation
(\ref{criticalcurrent}).  I believe the estimates in Table
\ref{table1} to be physically realistic because the total potential
difference across the sample $v_{\rm t}=rE_{\rm t}$ does not exceed
the maximum thermodynamic potential $\sqrt{2}e\Delta_0\sim$~0.7~meV
(where $\Delta_0\sim$~0.5~meV~\cite{harrison1}) that can be sustained
without destroying the CDW groundstate.

Next, I consider the coercion fields required to reverse the polarity
of the current.  On applying the simple expression for $H^\ast$ in
Equation (\ref{saturation}), the values estimated in Table
\ref{table1} compare favourably with experimental
observations~\cite{ardavan1}.  The behaviour of the irreversible
magnetization is also found to be very well described by a cubic law
of the form given by Equation (\ref{bean}).  Note that twice the
interval in field $2H^\ast$ is required to reverse the polarity of the
currents throughout the entire sample.  Meanwhile, the maximum
magnetic field that can be screened by the sample, $\chi_{\rm
irr}H^\ast$, is a small fraction of a millitesla.

Finally, I consider an odd situation that occurs at the minima and
maxima of the dHvA oscillations where the orbital susceptibility
$\chi=\partial M/\partial H$ vanishes.  Consequently, there will be
discrete fields at which $\chi_{\rm irr}\rightarrow$~0 causing the
irreversible magnetization never to reach saturation within the
quarter period of the dHvA oscillations $\Delta H=\mu_0H^2/4F$. 
Because there exists a chemical potential gradient $\nabla(\Delta\mu)$
across the sample, however, the situation where $M_{\rm irr}=0$ can
only exist in a narrow region of annular topology within the interior
of the sample.  The effect should still be observable, however, as
drop in the volume-averaged saturation magnetization.  This could
provide an explanation for the drop in the saturation magnetization
observed at intermediate filling factors at the lowest temperatures in
Reference~\cite{harrison1}.

In conclusion, I have identified a mechanism for persistent currents
in metals in which a CDW (or SDW) phase coexists with a strongly
field-dependent orbital magnetization.  The magnetic field drives an
exchange of carriers between the 1D and 2D Fermi surface sheets, which
has the effect of stretching or compressing the CDW (or SDW) like a
concertina.  Persistent currents result when pinning of the CDW (or
SDW) resists the concertina-like motions.  The charge polarization
field of the pinned CDW negates the Hall electric field that would
normally give rise to the Hall effect~\cite{chakraborty1}.  However,
because the CDW (or SDW) exists in a critical state on the brink of
sliding, the magnetic properties resemble those of a type II
superconductor~\cite{bean1}.  Like a superconductor, currents exist
only because of impurity pinning.  The mechanism for persistent
currents, described here, however, is neither the quantum Hall effect
or superconductivity, but one that is entirely different.  I propose
this effect as an explanation for persistent currents observed in
$\alpha$-(BEDT-TTF)$_2$KHg(SCN)$_4$~
\cite{harrison1,harrison2,harrison3,ardavan1}.

The work is supported by the Department of Energy, the National
Science Foundation (NSF) and the State of Florida.  I would like to
thank John Singleton, Albert Migliori and Arzhang Ardavan for
stimulating discussions.


\begin{figure}
\caption{Magnetic hysteretic behaviour anticipated by
the persistent current model for a cylindrical sample,
as described in the text.
}
\label{testloop}
\end{figure}

\begin{table}
\begin{tabular}{c|cccccccc}
filling & $\chi_{\rm mod}$ & $\chi_{\rm exp}$
& $M_{\rm max}$
& $j_{\rm c}$ & $E_{\rm t}$
& $v_{\rm t}$ & $\mu_0H^\ast$
& $\mu_0\chi_{\rm exp}H^\ast$ \\
~ & 10$^{-3}$ & 10$^{-3}$ & Am$^{-1}$ & Acm$^{-2}$ &
Vm$^{-1}$ & mV & T & mT \\ \hline
\hline 
half & -0.7 & -0.8 & 80 & 50 & 0.61
& 0.3 & 0.4 & 0.3 \\ \hline
int & -5 & -4 & 50 & 30 & 0.36
& 0.2 & 0.05 & 0.2

\end{tabular}
\caption{A comparison of the irreversible
susceptibility of $\alpha$-(BEDT-TTF)$_2$KHg(SCN)$_4$ expected for an
ideal sample according to the model, $\chi_{\rm mod}$, with that
obtained experimentally, $\chi_{\rm exp}$~[12].
Also listed 
are the maximum irreversible magnetization, $M_{\rm max}$, obtained
experimentally together with the critical current density, $j_{\rm c}$,
CDW depinning threshold electric field, $E_{\rm t}$, total potential
difference, $v_{\rm t}$, coercion field, $\mu_0H^\ast$, and screening
field, $\mu_0\chi_{\rm exp}H^\ast$, calculated according to the model.}
\label{table1}
\end{table}


\begin{references}
\bibitem{reason1} At magnetic fields too low for the dHvA
 effect to be realised, Landau diamagnetism is the only
 significant contribution to the orbital magnetization. For a
 2D closed pocket $M_{\rm Landau}\propto 1/g_{\rm 2D}$. The
 density of 2D states $g_{\rm 2D}$ is, if anything, only weakly
 dependent on $\Delta\mu$.
\end{references}
\end{document}